\documentclass{article}
\usepackage[utf8]{inputenc}
\usepackage{graphicx}

\title{Using Deep Learning to Improve Early Diagnosis of Pneumonia in Underdeveloped Countries
}
\author{Kyler Larsen}
\date{November 2021}

\begin{document}

\maketitle

\section{Introduction}
Increasing efficiency and accuracy for the diagnosis of pneumonia can decrease thousands of preventable deaths each year. Pneumonia is a common disease in the community that can be easily treated with inexpensive oral antibiotics administered by volunteer health care workers (Pneumonia 2021). However, if there is any delay in diagnosis or treatment, the disease course can be very serious and even deadly. Many of these diagnoses can be made through identification of symptoms such as cough, fever, or an increased respiratory rate. Chest x-ray is an inexpensive but vital component to diagnose pneumonia and to rule out other diseases. Certain types of pneumonia, especially atypical pneumonia such as Mycoplasma pneumoniae, can infect a person for weeks with no discernable symptoms. These types of pneumonia account for 10 to 40 percent of the cases of community acquired pneumonia in the United States (Atypical 2021). With this type of pneumonia it is even more important to use chest x-rays as an effective tool for accurate diagnosis. 

In underdeveloped countries, specifically in Sub-Saharan Africa and Southeast Asia, the death rate from pneumonia was almost 8-10 times the death rate in the United States in 2017 (Dadonaite et al 2021). Although many countries in Sub-Saharan Africa, such as Cameroon, receive medical equipment, including x-ray machines, through donation services, up to 70\% of the equipment is never used (Piaggio et al 2009). One primary cause of this problem is the lack of doctors and trained healthcare workers. In the Philippines, it was estimated that 85\% of Filipino nurses were working overseas in Western countries, leading to 70\% of Filipinos dying without medical attention (Finch 2013). In Sub-Saharan Africa, a critical shortage of doctors and nurses amounting to 2.4 million personnel in the years 2000-2006, hampered the ability of many health programs in place to detect and prevent chronic diseases (Naicker et al 2010). These statistics show that a shortage of trained medical personnel in underdeveloped countries causes delayed diagnoses or lack of proper treatment, leading to possible preventable deaths. This study will focus on evaluating a deep learning model, specifically over how well it can detect pneumonia in chest x-rays, and adopting the model to become an affordable and accurate alternative that will save many lives.

Previous studies have been conducted over the effectiveness of different AI models in limiting pneumonia’s harmful effects. For example, various machine learning models were evaluated on their accuracy in predicting severe pneumonia, which would allow proper treatment before the harmful effects could take place. The models evaluated in the study were the Support Vector Machine, Logistic Regression model, Random Forest Classifier, Naive Bayes, and AdaBoost. Ultimately, the study concluded that the Random Forest Classifier performed the best, due to its superior stability (Luo et al 2020). Another study evaluated a deep learning transfer model, VGG-16, as to how well it could identify pneumonia caused by COVID-19. The study also used chest x-ray analysis to form the model’s diagnoses, but concluded that further improvements could be made by adjusting hyperparameters and transfer learning combinations (M. D. Hasan et al 2021). This study attempts to fill this gap by evaluating a basic Convolutional Neural Network (CNN) by experimenting with hyperparameters to achieve maximum metrics.

The purpose of this study was to evaluate the performance of a CNN model given input in the form of chest x-rays. This study hypothesizes that the changing hyperparameters of the model can produce results equivalent to a trained physician. The null hypothesis assumes the model would not improve through training or would not achieve results similar to a physician. By testing various hyperparameters and extracting results based on key metrics, conclusions can be drawn as to how well the model performed (M. D. Hasan et al 2021). The significance of this study is to provide information regarding the feasibility of using a basic CNN model to diagnose accurately at point of care when there is shortage of trained healthcare workers. By developing and evaluating such a model, this study could save lives by improving efficiency and accuracy, as well as saving money and time.

\section{Methods}

In this study, the first step was to organize the data set. The data set used in this study is from Kaggle (Patel 2021), and features 6432 pre-classified images. This dataset includes 3 separate categories: COVID-19 pneumonia, general pneumonia, and normal lung x-rays. For this study, only images from the general pneumonia and normal lung x-rays were used, amounting to 2400 images with 1200 images from each classification. The data was then reshaped and augmented to provide a variety of data and to eliminate the need for “perfect” data. Augmentation included stretching certain pixels or rotating the image, allowing the creation of extra data as each image could have multiple augmentations done to it. The data was then split into 80\% for training data and 20\% for testing data. The training data is the data used by the model to learn its methods, while the testing data is used to evaluate the model’s accuracy. 

    A CNN works by receiving input in the form of an image. Then, through a hierarchical system, the various layers of the CNN create a network that resembles connected neurons, similar to the human brain. At each layer, an activation function transforms the results of each layer into specific output. The activation function used in this study is the Rectified Linear Activation Unit (ReLU), which returns the sum of each node’s input, or 0 if the sum is less than 0 (Brownlee 2021). The CNN then applies weights and flattens the output before it is fed to a feed-forward neural network and backpropagation is applied to every iteration of training. Over a series of these iterations, the model is able to distinguish between important dominant features and less-valuable low-level features (Brownlee 2021). 
    
    CNNs are valuable for identifying and classifying images, making it the ideal model for this study. In this study, the CNN was defined as sequential, since it only takes in one input and produces only one output, allowing for the addition of multiple layers. An initial layer of 32 filters with ReLU activation was then added, with a max pooling size of (2,2) to reduce output size. Then, additional layers with 64 filters were added with ReLU activation before the output was flattened. The amount of added layers is a tested variable, meaning it needs to be non-specific. Dropout requirements were also set at 0.5 in order to reduce overfitting, which is when the model memorizes the data set instead of forming its own predictions. In this case, 50\% of the neurons will be ignored during each iteration, forcing the model to learn new patterns each time (Brownlee 2021). Finally, the output is formatted and set to print loss function values, accuracy metrics, specificity, and sensitivity. 
    
    For the testing phase of the model, experimentation was done through changing the model’s hyperparameters. The first parameter experimented with was the number of extra layers added to the model, allowing more determinations to be made at the expense of overfitting. Other parameters include the number of epochs, or iterations, made by the model through the data set during each training cycle, and the set sensitivity of the model, allowing for the maximum sensitivity while still holding the other metrics constant. For each change to the parameters, 5 models were trained and the average values for each specificity, accuracy, precision, recall, and f1 score were taken to ensure accurate results. The output displays the accuracy and specificity of the model after each model is trained, as well as a graph depicting the trend of the accuracy over each epoch.

\section{Results}

The primary focus of the trials in this study is to maximize sensitivity while keeping specificity as high as possible. Sensitivity is defined by  \\
\begin{center}
\includegraphics{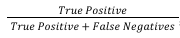} \\
\end{center}
meaning it determines how well the model identifies the disease. Specificity is defined by  \\ 
\begin{center} \includegraphics{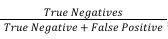} \end{center}  
which focuses on correctly identifying patients without the disease. The first trial increased sensitivity while monitoring specificity and accuracy, with the number of epochs set to 5 and the number of layers set to 3. From the graph shown in Figure 1 below, it was determined that the maximum sensitivity was around 85\%.
\begin{center}
\includegraphics[scale=0.75]{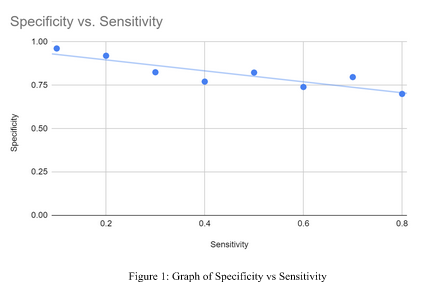} \\
\end{center}

The amount of extra layers was then adjusted, ultimately showing that the optimum sensitivity occurred with just 2 extra layers, resulting in a 90\% sensitivity with a 77.7\% specificity and an 82.5\% accuracy. To verify that these were the optimum values the model could produce, the amount of epochs was adjusted. 

\begin{center}
\includegraphics[scale=0.75]{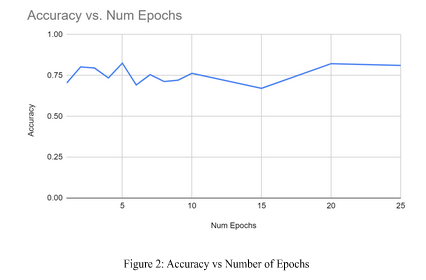} \\
\end{center}
\begin{center}
\includegraphics[scale=0.75]{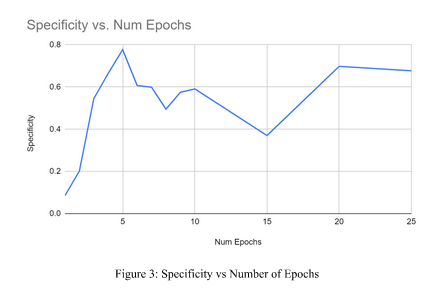} \\
\end{center}
From Figures 2 and 3, it was determined that 5 epochs was the optimum number of epochs, in terms of largest specificity and accuracy. This conclusion was further verified by its correlation to the f1-score of the test shown in Figure 4, which relies on both precision and recall.
\begin{center}
\includegraphics[scale=0.7]{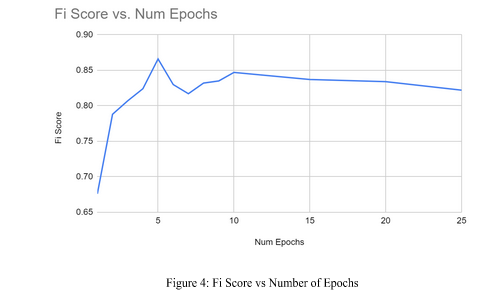} \\
\end{center}

\section{Discussion}
The main focus of this study was to evaluate the performance of a CNN model over an x-ray dataset, while maximizing sensitivity. The main reason for optimizing sensitivity was the nature of pneumonia and its treatments. Maximizing sensitivity focuses on eliminating false negatives at the cost of false positives, but ensures that almost all cases of pneumonia are identified. Maximizing specificity would focus on eliminating false positives at the expense of false negatives, meaning more cases of pneumonia would go undiagnosed. For studies similar to this one, the maximized metric will typically be positive predictive value, due to its measure of true positive accuracy. However, the goal of this model is to catch as many cases of pneumonia as possible, and maximizing the positive predictive value weighs both sensitivity and specificity equally. The treatment, especially if an early diagnosis is made, generally has minimal detrimental effects, meaning a false negative would have a more serious effect than a false positive. The primary treatment for bacterial pneumonia is antibiotics, which would likely have minimal negative effects on the person if taken with a false positive result (Pneumonia 2021). This means that while the rate of false negatives is important, the primary issue should be diagnosing all pneumonia positive images as positive rather than determining the accuracy of all positive diagnoses.
A primary issue faced in this study was the issue of overfitting. Overfitting happens when the model is trained too many times on the same data set, causing it to start memorizing the data set rather than learning from its attributes (Muralidhar 2021). Overfitting can be identified graphically and statistically. 

\begin{center}
\includegraphics[scale=0.75]{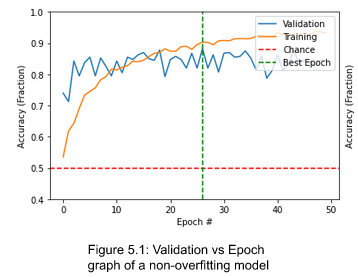}
\end{center}
\begin{center}
\includegraphics[scale=0.75]{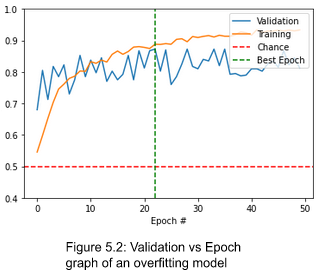}
\end{center}
Figure 5.1 shows when the model in the study has successfully learned a data set without overfitting, while Figure 5.2 shows when the model started overfitting. The first main differences are that the training line begins to plateau as it increases, which Figure 5.2 does to a greater degree than the left. Additionally, the training data’s accuracy surpasses the validation’s (test data) accuracy much sooner on Figure 5.2, illustrating a faster rate of learning which is common among overfitting models. This can also be identified numerically.
\begin{center}
\includegraphics[scale=0.75]{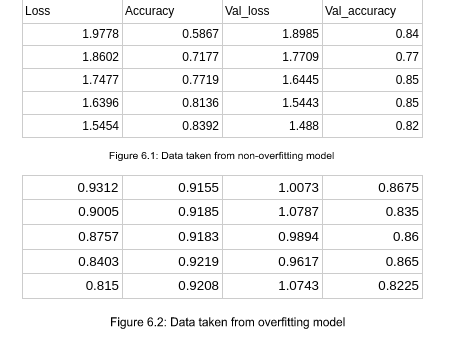}
\end{center}

Figure 6.1 shows results from the 5 epochs of the optimal CNN model, and Figure 6.2 shows the last 5 epochs from the CNN with training 20 epochs. Figure 6.1 shows a stagnant accuracy measurement, indicating the model is no longer learning, but rather memorizing, while Figure 6.2’s accuracy increases drastically. Additionally, the training accuracy for Figure 6.2 is much higher than the testing accuracy, which is also decreasing, indicating the model is not learning anymore, causing it to perform worse on the test sets. This demonstrates why a lower amount of epochs and training layers was more successful, as they prevented overfitting by reducing the amount of training. The danger with this is that it can cause underfitting, which happens when the model has not learned enough, causing bias towards the attributes it has learned. It is, therefore, important to test the model on as many values as possible. Overfitting was prevented by adding dropout layers and by lowering the amount of epochs and hidden layers of the model.

\section{Conclusion}
This study illustrates that the proposed model will be able to help those who cannot have proper  access to physicians for pneumonia diagnosis. The model achieved a sensitivity of 90\%, a specificity of 77\%, an accuracy of 82.5\%, and an f1 score of 0.866. These results are only passable, and overall provide more evidence to support the null hypothesis, because certainty is imperative when it comes to human lives. However, the results do indicate that the model has potential to save lives. The model was able to succeed in reducing overfitting, as well as preventing underfitting. A model like this one could be deployed as either an online or offline program able to receive x-rays in the form of images and return a positive or negative pneumonia diagnosis. No medical expertise other than an x-ray technician would be needed, allowing underserved areas to receive timely medical care, despite the shortage of medical personnel. Further research can explore the possibility of transfer learning as an alternative to the basic CNN, which could increase diagnosis accuracy.  Another important advancement could focus on applying this technology and methodology to other lung-based diseases that can be diagnosed through chest x-rays, such as tuberculosis. Therefore, with further research, the model can be improved so that underdeveloped countries without access to quality doctors or nurses can still receive the healthcare they need in a timely and affordable fashion.

\newpage

\section{Literature Cited}
ARASARATNAM, A. and HUMPHREYS, G. Emerging economies drive frugal innovation. World Health Organization. Bulletin of the World Health Organization, 2013. 91(1): p. 6. \\
\\
"Atypical (Walking) Pneumonia: Treatment \& Management". Cleveland Clinic, 2021, https://my.clevelandclinic.org/health/diseases/15744-pneumonia-atypical-walking-pneumonia.\\
\\
BROWNLEE, J. (2021). A Gentle Introduction to the Rectified Linear Unit (ReLU). Machine Learning Mastery. Retrieved 22 December 2021, from https:// \\machinelearningmastery.com/rectified-linear-activation-function-for-deep-learni\\ng-neural-networks/.\\
\\
"Causes Of Pneumonia | CDC". Cdc.Gov, 2021, https://www.cdc.gov/pneumo \\nia/causes.html.\\
CRISP, N., and CHEN, L. (2014). Global Supply of Health Professionals. New England Journal Of Medicine, 370(10), 950-957. https://doi.org/10.1056/nejmra\\1111610\\
\\
DADONAITE, B, and ROSER, M. "Pneumonia". Our World In Data, 2021, https://ourworldindata.org/pneumonia.\\
NAICKER S, EASTWOOD JB, PLANGE-RHULE J, TUTT RC. Shortage of healthcare workers in sub-Saharan Africa: a nephrological perspective. Clin Nephrol. 2010 Nov;74 Suppl 1:S129-33. doi: 10.5414/cnp74s129. PMID: 20979978.\\
\\
FINCH S. (2013). Philippines brain drain: fact or fiction?. CMAJ : Canadian Medical Association journal = journal de l'Association medicale canadienne, 185(12), E557–E558. https://doi.org/10.1503/cmaj.109-4459\\
LUO, Y., TANG, Z., HU, X., LU, S., MIAO, B., HONG, S., BAI, H., SUN, C., QIU, J., LIANG, H., \& NA, N. (2020). Machine learning for the prediction of severe pneumonia during posttransplant hospitalization in recipients of a deceased-donor kidney transplant. Annals Of Translational Medicine, 8(4), 82. doi:10.21037/atm.2020.01.09\\
\\
M. D. KAMRUL HASAN, SAKIL AHMED, Z. M. EKRAM ABDULLAH, MOHAMMAD MONIRUJJAMAN KHAN, DIVYA ANAND, AMAN SINGH, MOHAMMAD ALZAIN, MEHEDI MASUD, "Deep Learning Approaches for Detecting Pneumonia in COVID-19 Patients by Analyzing Chest X-Ray Images", Mathematical Problems in Engineering, vol. 2021, Article ID 9929274, 8 pages, 2021. https://doi.org/10.1155/2021/9929274\\
\\
MURALIDHAR, K. (2021). Learning Curve to identify Overfitting and Underfitting in Machine Learning. Medium. Retrieved 22 December 2021, from https://towardsdatascience.com/learning-curve-to-identify-overfitting-underfitt\\ing-problems-133177f38df5.\\ 
\\
PATEL, P. Chest X-ray (Covid-19 \& Pneumonia). Kaggle.com. (2021). Retrieved 22 December 2021, from https://www.kaggle.com/prashant268/chest-xray-covid19-pneumonia.\\
\\
PIAGGIO, D., MEDENOU, D., HOUESSUOVO, R., PECCHIA, L, Donation of medical devices in low-income countries: preliminary results from field studies. University of Warwick, Ecole Polytechnique d’Abomey-Calavi, University of Abomey-Calavi, 2009.\\
\\
"Pneumonia". Who.Int, 2021, https://www.who.int/news-room/fact-sheets/det\\ail/pneumonia.\\
\\
SAHA, S. (2021). A Comprehensive Guide to Convolutional Neural Networks\- the ELI5 way. towardsdatascience. Retrieved 22 December 2021, from https://to\\wards datascience.com/a-comprehensive-guide-to-convolutional-neural-networks-the-eli5-way-3bd2b1164a53.

\end{document}